\documentclass[a4paper,11pt]{article}
\usepackage{pdflscape}
\usepackage{amsmath}
\usepackage{cite}
\usepackage{amssymb}
\usepackage{dcolumn}
\usepackage{bm}
\usepackage{color}
\usepackage{epsfig}
\usepackage{amsfonts}
\usepackage{graphicx}
\usepackage{subfigure}
\usepackage{dcolumn}
\usepackage{indentfirst}
\usepackage{authblk}
\usepackage{slashed}
\usepackage{appendix}

\usepackage{booktabs}
\AtBeginDocument{
\heavyrulewidth=.08em
\lightrulewidth=.05em
\cmidrulewidth=.03em
\belowrulesep=.65ex
\belowbottomsep=0pt
\aboverulesep=.4ex
\abovetopsep=0pt
\cmidrulesep=\doublerulesep
\cmidrulekern=.5em
\defaultaddspace=.5em
}
\usepackage[colorlinks=false]{hyperref}
\hypersetup{citecolor=Blue}
\begin{document}

\vspace{80pt}
\centerline{\LARGE A comparison of condensate mass of QCD vacuum  between    }
\vspace{10pt}
{\LARGE \quad\quad\quad\quad Wilson line approach and Schwinger effect}

\vspace{40pt}

\centerline{
Sara Tahery,$^{a}$ 
\footnote{E-mail: s.tahery@impcas.ac.cn}
Xurong Chen $^{b}$ 
\footnote{E-mail:xchen@impcas.ac.cn}
Liping Zou $^{c}$
\footnote{E-mail:zoulp@impcas.ac.cn}}
\vspace{30pt}

{\centerline {$^{a,b,c}${\it Institute of Modern Physics, Chinese Academy of Sciences, Lanzhou 730000,  China
}}
\vspace{4pt}
{\centerline {$^{a,b,c}${\it University of Chinese Academy of Sciences, Beijing 100049, China
}} 
\vspace{4pt}
{\centerline {$^{b}${\it  Guangdong Provincial Key Laboratory of Nuclear Science, Institute of Quantum Matter,}}
{\centerline {$^{}${\it  South China Normal University, Guangzhou 510006, China}}

 \vspace{40pt}

\begin{abstract}
By  duality approach, we study condensate mass of QCD vacuum via dilaton wall background in presence of  parameter $c$  which
represents the  condensation in holographic set up. First from Wilson line calculation we  find $m_0^2$ (condensate parameter in mixed  non-local condensation)  whose behavior mimics that of QCD. The value of $m_0^2$ that we find by this approach, is in agreement with QCD data. In the second step we  consider produced mass $m$ via Schwinger effect mechanism in presence of parameter $c$.
  We show that generally vacuum condensation contribute mass dominantly and produced mass via Schwinger effect is suppressed by $m_0$ .
\end{abstract}

\newpage

\tableofcontents

%\newpage

%\maketitle
\section{Introduction}
Studying strong interactions in QCD, shows importance of vacuum condensate in both theory and phenomenology. Although the QCD sum rules is a basic tool in this issue, but a strong alternative is needed  for studying strongly coupled
gauge theories in a \textit{non-perturbative formulation}.

 AdS/CFT is a holographic description in which, a strongly coupled field theory on the boundary of the AdS space is mapped to the weakly coupled gravity theory in the bulk of AdS \cite{adscft,holo}. In this approach one starts from a five-dimensional effective
field theory somehow motivated by string theory and tries to fit it to QCD as much as possible. The first conjecture is based on conformal field theories, but  mass gap, confinement and supersymmetry breaking could be included \textit{by considering some modifications in gravity duals} \cite{apqcd,corr,qupo}.

A holographic model which represents the gluon condensation in a  gravity background with\textit{ Euclidean signature} is known with the following action \cite{action},
\begin{equation}\label{main action}
S=-\frac{1}{2k^2} \int d^5x \sqrt{g} \left(\mathcal{R}+ \frac{12}{L^2}-\frac{1}{2}\partial_{\mu}\phi \partial^{\mu}\phi\right). 
\end{equation}
In the above action $k$ denotes gravitational coupling in $5$-dimensions ,  $\mathcal{R}$ is Ricci scalar, $L$ is the radius of the asymptotic $AdS_5$ spacetime, and  $\phi$ is a massless scalar
which is coupled with the gluon operator on the boundary. To solve the Einstein equation and the dilaton
equation of motion, one needs a suitable ansatz which is the following dilaton-wall solution in \textit{Euclidean spacetime} \cite{action}, 
\begin{equation}\label{eq:metric}
ds^2= \frac{L^2}{z^2}(\sqrt{1-c^2 z^8}(dt^2+dx^2)+dz^2),
\end{equation} 
\begin{equation}\label{eq:dilaton}
\phi(z)=\sqrt{\frac{3}{2}} \log \dfrac{1+cz^4}{1-cz^4}+\phi_0,
\end{equation}
where $\phi_0$ is a constant, $c=\frac{1}{z^4_c}$  and $z_{c}$ denotes the IR cutoff. \footnote{ It shows that  $z$ is defined from zero to IR limit  as usual, so one should not misunderstand that  parameter $c$ bounds upper limit of $z$  to values less than cutoff ! in fact it should be interpreted as always $c<\frac{1}{z^4}$.}\\
  $x = x_1,x_2,x_3$ are orthogonal spatial
boundary coordinates. $z$ denotes the $5$th dimension, radial coordinate and $z=0$ sets the boundary. 
Expanding the dilaton profile near $z = 0$ will give,
\begin{equation}
\phi(z)=\phi_0+\sqrt{6} c z^4+...,
\end{equation}
where  parameter  $c$  characterizes
the QCD deconfinement transition,  with value $0< c\leq 0.9  GeV^4$ \cite{dico,cvalue,cvalue2}. \footnote{ In gauge/gravity dictionary, parameter $c$ in the metric background is related to the  condensation in the boundary theory.}
The dilaton field is dual to a scalar operator and the metric is dual to the energy-momentum tensor  of the dual field theory.
It is worth to mention that 
$\phi_0$  and c are
 the source and the parameter associated with
the color confinement scale respectively according to the holographic dictionary  \cite{source} (also for more discussion see \cite{dico2, hoqcd, tohoqcd}).
 The dilaton wall solution is related to the zero temperature case so it is  appropriate  to study condensate of \textit{vacuum} and its physics. 
 
 Another interesting phenomenon which could be studied in presence of  parameter $c$, is Schwinger effect. By definition pair production in presence of an external electric field is known as Schwinger effect in
non-perturbative QED \cite{SCH}. Due to this phenomenon when the external field is strong enough the virtual electron-positron pair become real particles. In other words vacuum is destroyed in presence of such a field.
Although this context had been considered in QED first, it is not restricted to it any more and it has been extended to QCD \cite{uah}.
According to this mechanism, the vacuum decay can be considered in presence of a deformation parameter \cite{vacsc}, also  potential analysis of Schwinger effect in presence of  condensation parameter has been done in \cite{glusch}. Therefore we skip this part and interested reader can refer to the mentioned references. But produced mass in such a Schwinger effect mechanism is one of our interests in this study.

This paper is organized as follows. In section \ref{se:Mass from gluon condensate} we will follow Wilson line calculation method based on \cite{GCAn} to discuss condensation.
 In section \ref{sec: Mass from Schwinger effect} by considering standard form of brane embedding in Schwinger effect holographic set up, we will calculate $m$ as produced mass  where the effect of  condensation parameter will be taken into account. Having results from both above approaches we will compare them. In section \ref{sec:Conclusions} conclusions are given.
\section{Mass from  condensation}\label{se:Mass from gluon condensate}
Before starting our calculation we review method of \cite{GCAn}, so if the reader is familiar with this reference can skip this part and continue from \eqref{eq:string action}.
Condensation is associated with a five dimensional operator
constructed from the quark ($q \bar{q} $) and gluon field ($G_{mn}$) as,
\begin{equation}\label{eq:expGC}
<g \bar{q}\sigma^{mn} G_{mn}q >=m_0^2<\bar{q}q>,
\end{equation}
where  $<\bar{q}q>$ shows quark condensation, $\sigma^{mn}$ is an antisymmetric combination of matrices and $g$ is a gauge coupling constant. \textit{Then, $m_0^2$ appears as a constant of proportionality in the parametrization}. In fact in analysis of condensation by QCD sum rules $m_0^2$ appears in relation between the zeroth moment of quark-gluon distribution function and the first moment of distribution function of quarks in the vacuum (see \cite{piwa} for exact discussion). Here we avoid that calculation and follow holographic set up while we keep in our mind that $m_0^2$ is not mass of hadrons but is a coefficient (with $(mass)^2$ dimension) of quark condensate in the simplest quark-gluon mixed condensate.  Let us consider a non-perturbative gauge invariant correlator as,
\begin{equation}\label{eq:corr}
\Psi(x_a,x_b)=<\bar{q}U_P (x_a,x_b)q>,
\end{equation}
where $U_P (x_a,x_b)$ is a path-ordered Wilson line defined as,
\begin{equation}\label{eq: path-ordered Wilson line}
U_P (x_a,x_b) = P \exp[ig\int_{0}^{1} dl \frac{dx^{\mu}}{dl}A_{\mu}(x(l))],
\end{equation}
where $\mu$ is running over the 4 dimensional indices and $l$ is a parameter of the path running from $0$ at $x = x_a$ to $1$ at $x = x_b$. The path is taken to be a straight line. If one sets \cite{piwa},
\begin{equation}\label{eq:Psi}
\Psi(x_a,x_b)=<\bar{q}q> Q(r),
\end{equation}
then $m^2_0$ is given by the coefficient of $r^2$, with $r=|x_a-x_b|$, in the expansion of the function $Q$ as $r\longrightarrow 0$,
\begin{equation}\label{Q}
Q(r) =1-\frac{1}{16}m_0^2 r^2+ \mathcal{O}(r^4),
\end{equation}
which holds in\textit{ Euclidean space} and in Minkowski space it is modified by $r^2\longrightarrow -r^2$.\\
 Author of the mentioned reference set an ansatz for computing the function $Q$ within gauge/string duality.
  The quark operators,  the Wilson line on a four-manifold (which is the boundary
of a 5-dimensional manifold) and the function $Q$ are given in terms of the area
(in string units) of a surface in the 5-dimensional manifold by
\begin{equation}\label{Q ansatz}
Q(r) =e^{-S}.
\end{equation}
\begin{figure}[h!]
\begin{center}$
\begin{array}{cccc}
\includegraphics[width=100 mm]{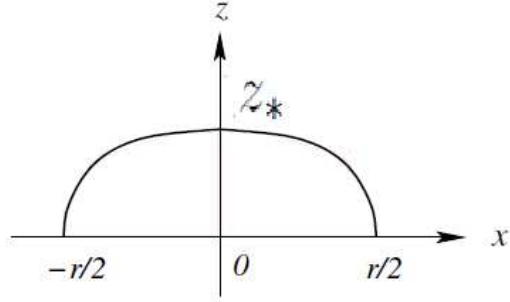}
\end{array}$
\end{center}
\caption{Surface of the 5- dimensional manifold, with boundary at $z=0$. The curved profile of the static string is stretched between quark sources from $-\frac{r}{2}$ and  $\frac{r}{2}$,  the surface is bounded between this and straight Wilson line along x-axis, $z_\ast$ is turning point of the string.}
\end{figure}\label{fig:fig}
Now let us back to our case. After warming up with above review, it is understandable if we find $Q$ as a function of $r$ then we can approximate $m_0^2$ parameter.

In continue we will find shape of the string describing the quark source. Let's consider the Nambu-Goto action,
 \begin{equation}\label{eq:string action}
S=\frac{1}{2\pi\alpha'} \int d\sigma^2 e^{\frac{\phi}{2}} \sqrt{\det \mathcal{G}_{\mu\nu} \partial_{\alpha} X^{\mu} \partial_{\beta} X^{\nu} },\quad  \quad\quad \quad \quad g_{\alpha\beta}= \mathcal{G}_{\mu\nu} \partial_{\alpha} X^{\mu} \partial_{\beta} X^{\nu}
\end{equation}
where $\sigma^{\alpha}$ are coordinates on the string worldsheet, $\mathcal{G}_{\mu\nu}$ is the space-time metric \eqref{eq:metric}, $g_{\alpha\beta}$ is induced metric, $\alpha'$ is universal Regge slope and the embedding
of the string worldsheet in spacetime is,
\begin{equation}\label{eq:Xsigma}
X^{\mu}(\sigma)=(t,X_1,X_2,X_3,z). 
\end{equation} 
%\newpage
Then from the background metric \eqref{eq:metric}, dilaton field \eqref{eq:dilaton} 
and in static gauge $\sigma_{1}=t,\quad 0<t<\tau$ , $\sigma_{2}=X_1=x, \quad -\frac{r}{2}<x< \frac{r}{2}$  one may obtain\footnote{In this calculation we wrote
$\exp  \frac{\phi(z)}{2}= \exp \log (\frac{1+cz^4}{1-cz^4})^{\sqrt{\frac{3}{8}}}= (\frac{1+cz^4}{1-cz^4})^\frac{\sqrt{\frac{3}{8}}}{\ln10}= (\frac{1+cz^4}{1-cz^4})^{0.266}$, (recall that $\log A= \frac{\ln A}{\ln 10}$ and $\exp \frac{\ln A}{\ln 10}=A^{\frac{1}{\ln 10}}$) for this reason our \eqref{eq:static gauge action} seems different with that of \cite{cvalue}   while they are same. Also in our calculation 
 only \textbf{the first digit after the decimal point} is taken into account so  $(1+cz^4)^ {0.516}\approx \sqrt{1+cz^4}$ . In addition one can set $\phi_0=0$ in \eqref{eq:dilaton} according to \cite{cvalue,cvalue2}.}\label{foot:action}.

\begin{equation}\label{eq:static gauge action}
S=\frac{1}{2\pi\alpha'} \int_{0}^{\tau} dt \int_{-\frac{r}{2}}^{\frac{r}{2}} dx\quad\frac{\sqrt{1+cz^4}}{z^2} \left(\sqrt{1-c^2z^8}+(\frac{dz}{dx})^2\right)^\frac{1}{2}, 
\end{equation} 
where we set the radius of spacetime $L=1$. Then the following relation should be satisfied,  
\begin{equation}\label{eq:eq of mo}
\frac{\partial \mathcal{L}}{\partial(\partial_x z)}\partial_x z-\mathcal{L}= Const.
\end{equation}
 And from the boundary condition\footnote{From the string configuration it is easy to find that  $z_*$, turning point of the string shows maximum of $z$. },
\begin{equation}\label{eq:bouncon}
at \quad z=z_{*}, \quad\Rightarrow \quad\frac{dz}{dx}=0,
\end{equation}
the constant value in that right hand side of \eqref{eq:eq of mo} could be found. Thus the solution of \eqref{eq:eq of mo} is,
\begin{equation}\label{eq:shape of string}
\frac{dx}{dz}=\frac{1}{\sqrt{(\dfrac{z_*}{z})^4 \frac{(1+c z^4)(1-c^2 z^8)}{(1+c z_*^4)\sqrt{1-c^2z_*^8}}-\sqrt{1-c^2z^8}}}.
\end{equation}
With change of variable $u=\dfrac{z}{z_*}$,  and using the boundary conditions from Fig.1, as 
\begin{eqnarray}
x=0\Longrightarrow z=z_*\nonumber\\
x=\pm\frac{r}{2}\Longrightarrow z=0,
\end{eqnarray}
   shape of the string appears as,
\begin{equation}\label{eq:randu1}
x(z)=z_*\int_{u}^{1} \dfrac{ u^2 du}{\sqrt{\frac{(1+c z_*^4 u^4)(1-c^2 z_*^8 u^8)}{(1+c z_*^4)\sqrt{1-c^2z_*^8}}-u^4\sqrt{1-c^2z_*^8 u^8}}}.
\end{equation}   
Therefore at $z=0$ and $-\frac{r}{2}<x<\frac{r}{2}$, \eqref{eq:randu1}  becomes,
   \begin{equation}\label{eq:randu}
r=z_*\int_{0}^{1} \dfrac{ u^2 du}{\sqrt{\frac{(1+c z_*^4 u^4)(1-c^2 z_*^8 u^8)}{(1+c z_*^4)\sqrt{1-c^2z_*^8}}-u^4\sqrt{1-c^2z_*^8 u^8}}},
\end{equation}
which is length of the string. After describing the string shape, now we need to calculate the renormalized area of the surface in figure 1. \\
So we choose the gauge $\sigma_{1}=X_1=x$ and $\sigma_{2}=z$, then from the action \eqref{eq:string action} and string profile \eqref{eq:randu1} we lead to \footnote{Same approximations with what have been done in calculation of \eqref{eq:static gauge action}.},
\begin{equation}\label{eq:choosed gauge action}
S=\frac{1}{2\pi\alpha'}  x(z) \int_{0}^{z_*}dz \quad\frac{\sqrt{1+cz^4}}{z^2} . 
\end{equation}
 To regularized the above integral one cut off the integral
at $z=\epsilon$. Therefore from (\ref{eq:choosed gauge action}) the regularized answer of the integral is,
\begin{equation}\label{eq:regularizedS}
S_{reg}=\frac{r}{2\pi \alpha' z_{*}}(\frac{1}{\epsilon}-\quad_{2}F_1  [-\dfrac{1}{2},-\frac{1}{4},\frac{3}{4}, -cz_*^4]).
\end{equation}
 $_{2}F_1$ is hypergeometric function and we subtract $(\frac{r}{2\pi \alpha' z_{*}}(\frac{1}{\epsilon}-a))$ to deal with the power divergence where $a$ is a constant must be specified from
renormalization conditions, so the answer of the integral (\ref{eq:choosed gauge action}) is  given by the hypergeometric function as,
\begin{equation}\label{eq:regularizedS2}
S_{reg}=a-\frac{r}{2\pi \alpha' z_{*}}\quad_{2}F_1 [-\dfrac{1}{2},-\frac{1}{4},\frac{3}{4}, -cz_*^4].
\end{equation}
Up to now, we have found both $r$ \eqref{eq:randu} and the action  \eqref{eq:regularizedS2} describing the renormalized area. Relating them to each other, we will have behavior of  function $Q$ versus $r$. It is not clear how to  do this but by considering two important limiting cases, long distances and short distances, we can analyze the behavior of the function.\\
We begin with \eqref{eq:randu}. By considering $c_1=cz_*^4$, one can check $c_1\longrightarrow 0$ leads to long $r$ and $c_1\longrightarrow 1$ leads to short $r$.
Expanding right hand side of (\ref{eq:randu}) around these two values of $c_1$ will show behavior of $r$ at limiting cases.
So at short distances (we will denote by sd)   the asymptotic behavior of (\ref{eq:randu}) is given by mathematical gamma function \footnote{Remember that gamma function satisfies, $\Gamma(y+1)=y \Gamma(y)$ and $\Gamma(\frac{1}{2})=\sqrt{\pi}$. } as,
\begin{equation}\label{eq:short r}
r_{sd}\approx \sqrt[4]{c^3}z_*^4,
\end{equation}
and at long distances (we will denote by ld)  the asymptotic behavior of (\ref{eq:randu}) is given by,
\begin{equation}\label{eq:long r}
r_{ld}\approx \dfrac{\sqrt{\pi}}{6} z_*(2+cz_*^4)\frac{ \Gamma(\dfrac{7}{4})}{ \Gamma(\dfrac{5}{4})}.
\end{equation}

In the same manner we expand (\ref{eq:regularizedS2}) around  $c_1\longrightarrow 0$  and $c_1\longrightarrow 1$ to find behavior of the action at long distances and short distances respectively.
 Then at short distances, the action (\ref{eq:regularizedS2}) behaves as,
\begin{equation}\label{eq:short distance action}
S_{sd}=\frac{r}{2\pi \alpha' z_{*}}(-1+\frac{\sqrt{\pi}}{8}\frac{ \Gamma(\dfrac{3}{4})}{ \Gamma(\dfrac{5}{4})} cz_*^4)+a,
\end{equation}
and at long distances, the action (\ref{eq:regularizedS2}) behaves as,
\begin{equation}\label{eq:long distance action}
S_{ld}=\frac{r}{2\pi \alpha' z_{*}}(-1+\frac{cz_*^4}{6})+a.
\end{equation}
First we should fix the value of $a$. According to the standard normalization of $Q$ we may impose the condition $Q(0) = 1$ \cite{piwa}, which gives $a=0$.\\
Combining (\ref{eq:short distance action}) with (\ref{eq:short r}) we find the desired behavior of the function $Q$ at short distances as,\\
\begin{equation}\label{eq:Q short}
Q=1-\frac{1}{2\pi \alpha'}(\frac{3}{10})\sqrt{c}R^2+\mathcal{O}(r^4),
\end{equation}
where $R=r-\varepsilon$ and $\varepsilon$ is a  small constant value. Finally comparing \eqref{Q}  with \eqref{eq:Q short} we may approximate mass at short distances as,
\begin{equation}\label{eq:short mass}
m_0^2\approx\frac{12}{5}\frac{\sqrt{c}}{\pi \alpha'}.
\end{equation}
Considering $\frac{1}{\alpha'}=0.94$ and $c=0.9 GeV^4$ (from the fit to the slope of the Regge trajectories \cite{corr}), we can estimate $m_0^2$ as,
\begin{equation}\label{eq:short mass value}
m_0^2\approx 0.63 GeV^2.
\end{equation}
According to the original
phenomenological estimate based on the QCD sum rules \cite{qcdm} it is given
by $m^2_0=0.8 \pm 0.2 GeV^2$, therefore \eqref{eq:short mass value} shows an acceptable result.
 Notice that the  condensation we are  discussing in this paper  is a mixed condensate of quarks and gluons \eqref{eq:expGC} but not exactly the gluon condensate $\alpha_s G^2$  values which in 
 different literature  have been extracted from phenomenology of QCD  \cite{npb147,plb387, LTH 668, ol070,st110,yu091,re190},  mostly in range of $0.024 GeV^4$- $0.07 GeV^4$. 

To deal with long distances we combine (\ref{eq:long distance action}) with (\ref{eq:long r}) and consider leading order of $r$, shows that at long distances the function $Q$ decays exponentially, as it is expected in QCD. 
\begin{eqnarray}\label{long decay}
Q&=&e^{-S}, \nonumber\\
&=& e^{-M^2 r^2},
\end{eqnarray}
where,
\begin{equation}\label{eq:long mass value}
M\approx \dfrac{1}{\sqrt{12\pi \alpha'}}\sqrt[4]{c}.
\end{equation}
In fact in $4d$ QCD \textit{linear} decay is expected, while \eqref{long decay} shows a faster decay according to \textit{quadratic} term $r^2$. It comes from the nature of the $GeV^4$ modification of background metric, which represents  condensation in gravity dual. So such a dimension of $c$ leads to exponential function of $Q$ as $-M^2 r^2$ rather than that of ref \cite{GCAn} which is $-Mr$ in presence of modification parameter with $GeV^2$ dimension. Notice that both dimensions of results are the same.  To check availability of this result,
let us estimate the above $M$, with  $\frac{1}{\alpha'}=0.94$ and $c=0.9 GeV^4$, it is given by,
\begin{equation}\label{eq:long mass value2}
M\approx 0.15 GeV,
\end{equation}
whose value is close to pion mass. Same with result of \cite{GCAn}, it is understood that at long distances the correlator is dominated by the lightest meson contribution.
 Having found an estimate for function $Q$ at  short distances and behavior of that at long distances, we close this section and in the next step we will study produced mass by  Schwinger effect in presence of  condensation parameter.
\section{Mass from Schwinger effect}\label{sec: Mass from Schwinger effect}
Recall that Schwinger effect is about vacuum decay in presence of an external electric field. In fact before vacuum decay, there is a potential  barrier as $V_{CP+SE}$ which is the sum of the Coulomb potential (CP) and static
energy (SE). Once the external field is turned on, the total potential is $V_{tot}=V_{CP+SE}-Ex$  that includes electrostatic potential between test particles (the virtual particles in the vacuum which become real after decay of vacuum). In this relation $E$ is the external electric field and $x$ is the distance between particles. It is clear that increasing field will suppress the potential barrier after a critical value, so vacuum decays and virtual test particles become real \cite{pah,shco,shco2}. \\

\begin{figure}[h!]
\begin{center}$
\begin{array}{cccc}
\includegraphics[width=100 mm]{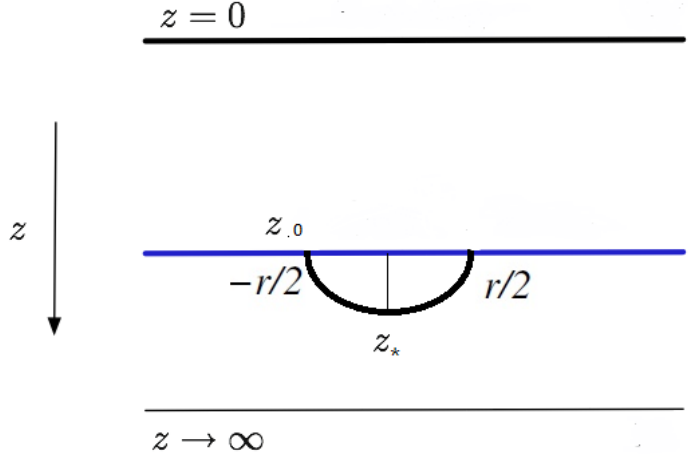}
\end{array}$
\end{center}
\caption{The holographic set up to consider Schwinger effect  test particles.}
\end{figure}
As we mentioned before, the Schwinger effect  potential analysis in presence of gluon condensation has been studied in \cite{glusch}. In this section we follow mass production in the same mechanism.\\
Consider a pair of virtual particles in vacuum which will be real particles in presence of an \textit{strong enough} electric filed. This procedure  corresponds to  holographic potential barrier decay  by gauge/gravity duality. If one considers the produced pair in the bulk then the mass value goes to infinity,   to avoid the divergent mass in a holographic way a probe D3-brane attached to the test particles is located at an \textit{intermediate position} $z_0$ \cite{seza}. 
Then the mass formula is given by,
\begin{equation}\label{Schmass formula}
m=\frac{1}{2\pi\alpha'} \int_{z_0}^{\infty} dz \quad  e^{\frac{\phi}{2}} \sqrt{\det \mathcal{G}_{\mu\nu} \partial_{\alpha} X^{\mu} \partial_{\beta} X^{\nu} },
\end{equation}
thus with the induced metric on the string world-sheet  and in static gauge  one may obtain,
\begin{equation}\label{eq:Schwinger mass1}
m=\frac{1}{2\pi\alpha'} \int_{z_0}^{\infty}  dz\quad\frac{\sqrt{1+cz^4}}{z^2} \left(\sqrt{1-c^2z^8}+(\frac{dz}{dx})^2\right)^\frac{1}{2}. 
\end{equation} 
Combining this with (\ref{eq:shape of string}) one may find,
\begin{equation}\label{eq:Schwinger mass2}
m=\frac{1}{2\pi\alpha'} \int_{z_0}^{\infty}  dz\quad\frac{\sqrt{1+cz^4}}{z^2}{\sqrt{(\dfrac{z_*}{z})^4 \frac{(1+c z^4)(1-c^2 z^8)}{(1+c z_*^4)\sqrt{1-c^2z_*^8}}}} . 
\end{equation}
With changing variable $y=\frac{z}{z_0}$ and $b=\frac{z_0}{z_*}$, (\ref{eq:Schwinger mass2}) is,
\begin{equation}\label{eq:mass with y}
m=\frac{1}{2\pi\alpha'}\frac{1}{b^2 z_0} \frac{1}{\sqrt{(1+c_1)\sqrt{(1-c_1^2)}}}\int_{1}^{\infty} dy\quad y^{-4} (1+c_1 b^4 y^4)^{\frac{3}{2}}(1-c_1 b^4 y^4)^{\frac{1}{2}},
\end{equation}
recall that we defined $c_1=c z_*^4$.  Integral \eqref{eq:mass with y} is solvable analytically\footnote{Notice that this integral has been solved analytically by mathematica without any approximation, after finding a very big result with many terms according to the fact that always  $b<1$  we kept the leading term  and ignored many terms with 8, 16,... power of $b$.}, however since $b<1$ we consider the leading term of $b$  in the answer and ignore others. Then the answer is,
\begin{equation}\label{eq:mass answer}
m=\frac{1}{6\pi \alpha' z_0 b^2}\left( \frac{1}{\sqrt{(1+c_1)\sqrt{(1-c_1^2)}}}\right).
\end{equation}
We expand \eqref{eq:mass answer} near  $c_1\longrightarrow 0$ and $c_1\longrightarrow 1$ to consider mass at long distances and short distances respectively.
 At long distances (\ref{eq:mass answer}) behaves as, 
\begin{equation}\label{eq:Schwinger mass at long distance}
m=\frac{1}{6\pi \alpha' z_0 b^2}(1-\frac{c_1}{2})+\mathcal{O}(c_1^2),
\end{equation}
and at short distances, 
\begin{equation}\label{eq:Schwinger mass at short distance}
m=\frac{1}{6\sqrt[4]{2}\pi \alpha' z_0 b^2} \frac{1}{\sqrt{c_1^3}}.
\end{equation}
 Combining (\ref{eq:Schwinger mass at short distance}) with (\ref{eq:short r}) and (\ref{eq:Schwinger mass at long distance}) with (\ref{eq:long r}) we may study mass at short distances and long distances respectively.
 At long distances  produced mass in Schwinger effect  behaves as, 
\begin{equation}\label{eq:Schwinger mass at long distance2}
m=\frac{1}{6\pi \alpha' z_0 b^2}(1-\frac{\sqrt[5]{(c r^4)}}{2}),
\end{equation} 
which shows with increasing $r$ mass goes to zero.
 And at short distances mass is given by,
\begin{equation}\label{eq:Schwinger mass at short distance2}
m=\frac{1}{6\sqrt[4]{2}\pi \alpha' z_0 b^2} \frac{1}{\sqrt[8]{(c r^4)^3}},
\end{equation}
according to which, any discussion on produced mass in Schwinger effect significantly depends on brane position $z_0$ in the bulk,  (in fact since $b=\frac{z_0}{z_*}$, having one of $b$ or $z_0$  is enough). So, although we have phenomenological value for parameter $c$ which can relate  distance of produced pair and mass,  $z_0 b^2$ always needs to be considered as some coefficient. 

After deriving mass formula in Schwinger effect in presence of  condensation , it is interesting to compare the result with that of previous section.
 Needless to say that what we have found in last section is very different in its nature with Schwinger effect produced mass. $m_0^2 $ which appears in non-local and mixed condensates  is a constant of proportionality in the conventional parametrization \footnote{Notice that even considering $m_0$ as exact value of quark mass could not be correct , as we mentioned in last section the value of this parameter has been estimated  by  original
phenomenological estimate based on the QCD sum rules \cite{qcdm}, it is given
by $m^2_0=0.8 \pm 0.2 GeV^2$. }. What we have up to now, are two different values by two different approaches. \textit{$m$ from Schwinger effect is exactly mass of quark antiquark which have been produced by vacuum decay}. Although the responsible for vacuum decay and pair production is the external field $E$,\textit{ but in the background there is parameter} $c$. On one hand\textit{ if one accepts that in any mass production from vacuum some kind of  condensation appears, then  existence of parameter $c$ in holographic Schwinger effect makes sense. On the other hand we have $m_0^2$ from Wilson line calculation directly, means the condensation is happening based on $c$}. So these two approaches are different in what is responsible for mass production also in string and/ or brane configuration which leads to different computations, but\textit{ they have parameter $c$ which plays important role in both}.
Above explanations motivate us to compare $m$ and $m_0$. 
 Considering both (\ref{eq:Schwinger mass at short distance}) and (\ref{eq:short mass}) we can find the following ratio approximately, 
\begin{equation}\label{eq:mass fraction}
\frac{m_0}{m}\approx 20 z_0 b^2 r^2 \sqrt[4]{c^3}.
\end{equation}
As an example lets consider position of probe brane in the bulk as $z_0=\frac{z_*}{2}$, knowing the mentioned position we can find the fraction as,
\begin{equation}\label{eq:mass middle fraction}
\frac{m_0}{m}\approx \frac{9}{4} \sqrt[16]{(c r^4)^9}.
\end{equation}
If we consider the intermediate position of the embedded brane in the bulk, then it is interesting to study the behavior of $\frac{m_0}{m}$ schematically.\\
\begin{figure}[h!]
\begin{center}$
\begin{array}{cccc}
\includegraphics[width=100 mm]{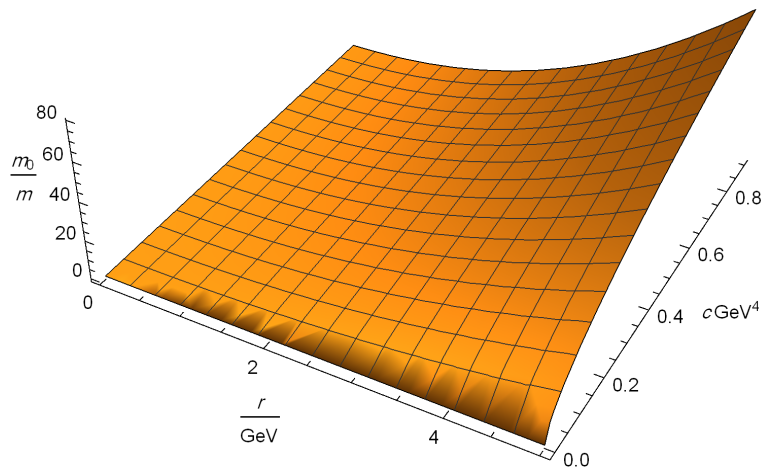}
\end{array}$
\end{center}
\caption{Behavior of $\frac{m_0}{m}$ versus distance $(GeV^{-1})$ and parameter $c (GeV^4)$.}
\end{figure}\label{fig:massrc}
Figure \ref{fig:massrc} shows $\frac{m_0}{m}$ as a function of distance and $c$. The distance has been considered $0< r< 1 fm$ or $5 (GeV)^{-1}$, also $c$ axis denotes value $0< c\leq 0.9  GeV^4$. Considering parameter $c$ around our desired value $0.9 GeV^4$ with distance  $r\simeq 0.14 fm= 0.7 GeV^{-1} $ lead to $\frac{m_0}{m}=1$.  Increasing $r$ after this crucial value, increases the  ratio $\frac{m_0}{m}$ significantly. 
Notice that $r$ in string configuration corresponds to  diameter of meson. As a case, lets consider the ratio $\frac{m_0}{m}=1$ and $r\approx 2.6 GeV^{-1}=0.52 fm$ corresponds to $J/ \psi$ diameter. It leads to  very small value of  the parameter $c\approx 0.005 GeV^4$.  This value of $c$   gives us the near zero value of $m_0^2\approx 0.003 GeV^2$ from \eqref{eq:short mass}. In comparison with \eqref{eq:short mass value} it is far from  acceptable value which fit QCD data. But   $J/ \psi$ with $c\approx0.9  GeV^4$ results in  $\frac{m_0}{m}\approx 18$. Therefore  condensation of vacuum suppresses Schwinger effect mass obviously. As a physics interpretation, one may relate that result. To the best of our knowledge Schwinger effect has not produced mass in experiment yet. Although the theory is clear and comprehensive enough (from QED to QCD and even higher dimensional objects) but according to the nature of this phenomenon it is not easy to reach the energy level which could destroy vacuum. On the other hand  condensation (of gluons, quarks,...) does happen in hadrons structure and all QCD structure is contained within hadrons, sounds it is universal in QCD context. From this  view it makes sense that Schwinger effect mass is suppressed  by  condensation of vacuum.
\section{Conclusions}\label{sec:Conclusions}
Gluon condensation is an important issue since it represents many phenomenological aspects of QCD. In this work we considered  dilaton wall background  related to zero temperature to study vacuum condensation. First we calculated $m^2_0$  by Wilson line calculation. Our results satisfy QCD data at both short distances and long distances. In the second step we studied another  mass production mechanism, Schwinger effect in presence of gluon condensation in which an external electric field is responsible for vacuum decay. These two mechanisms naturally are different in approaches and physics both. However \textit{since gluon condensation  plays role in both of them}, it was our interest to compare their results. We ended up by finding  ratio of $\frac{m_0}{m}$ as a function of $r$. We have found that generally the produced mass via Schwinger effect  is suppressed by condensation.\\
\\
\textbf{Acknowledgement}\\
This work was supported by the National Natural Science Foundation of China(Grant No. 11575254), the National Key Research and Development Program of China 
(No. 2016YFE0130800), 
and the Strategic Priority Research Program  of Chinese  Academy of Sciences (Grant No. XDB34030301). ST is supported by the PIFI (grant No. 2021PM0065).

\end{document}